\documentclass[conference,10pt]{IEEEtran}
\IEEEoverridecommandlockouts

\usepackage{tabularx}

\usepackage{tikz}
\usetikzlibrary{shapes.geometric}
\usetikzlibrary{backgrounds}

\usepackage{cite}
\usepackage{amsmath,amssymb,amsfonts, amsthm}
\usepackage{graphicx, booktabs}
\usepackage{textcomp}
\usepackage{amsmath,amsfonts}
\usepackage[linesnumbered, vlined, ruled]{algorithm2e}
\usepackage[noabbrev,capitalise,nameinlink]{cleveref}
\usepackage{xcolor, colortbl}
\usepackage{circuitikz}
\usepackage[a4paper, total={184mm,239mm}]{geometry}
\usepackage{balance}
\usepackage{todonotes}
\usepackage{array}
\usepackage[export]{adjustbox}
\usepackage{stfloats}
\usepackage{booktabs}
\usepackage{makecell}
\usepackage{xspace}
\usepackage{siunitx}
\usepackage[inline]{enumitem}
\usetikzlibrary{calc}
\usepackage{balance}

\newtheorem{theorem}{Theorem}

\usepackage{tikz-3dplot}
\Crefname{figure}{Fig.}{Figs.}
\def\BibTeX{{\rm B\kern-.05em{\sc i\kern-.025em b}\kern-.08em
    T\kern-.1667em\lower.7ex\hbox{E}\kern-.125emX}}

\newcolumntype{L}[1]{>{\raggedright\let\newline\\\arraybackslash\hspace{0pt}}m{#1}}
\newcolumntype{C}[1]{>{\centering\let\newline\\\arraybackslash\hspace{0pt}}m{#1}}
\newcolumntype{R}[1]{>{\raggedleft\let\newline\\\arraybackslash\hspace{0pt}}m{#1}}

\begin{document}

\title{Scalable Sequential Optimization Under Observability Don't Cares\\}

\author{\IEEEauthorblockN{Dewmini Sudara Marakkalage\IEEEauthorrefmark{1},
Eleonora Testa\IEEEauthorrefmark{2}, 
Walter Lau Neto\IEEEauthorrefmark{2},
Alan Mishchenko\IEEEauthorrefmark{3}, \\
Giovanni De Micheli\IEEEauthorrefmark{1},
Luca Amarù\IEEEauthorrefmark{2}}
\IEEEauthorrefmark{1}Integrated Systems Laboratory, EPFL, Lausanne, Switzerland\\
\IEEEauthorrefmark{2}Synopsys Inc., Design Group, Sunnyvale, California, USA\\
\IEEEauthorrefmark{3}Department of EECS, University of California, Berkeley, USA}

\maketitle

\begin{abstract}

Sequential logic synthesis can provide better Power-Performance-Area~(PPA) than combinational logic synthesis since it explores a larger solution space. 
As the gate cost in advanced technologies keeps rising, sequential logic synthesis provides a powerful alternative that is gaining momentum in the EDA community.
In this work, we present a new scalable algorithm for don’t-care-based sequential logic synthesis. Our new approach is based on sequential $k$-step induction and can apply both redundancy removal and resubstitution transformations under Sequential Observability Don’t Cares~(SODCs). Using SODC-based optimizations with induction is a challenging problem due to dependencies and alignment of don’t cares among the base case and the inductive case. We propose a new approach utilizing the full power of SODCs without limiting the solution space.
Our algorithm is implemented as part of an industrial tool and achieves 6.9\% average area improvement after technology mapping when compared to state-of-the-art sequential synthesis methods. Moreover, all the new sequential optimizations can be verified using state-of-the-art sequential verification tools. 
\end{abstract}


\section{Introduction}
\label{sec:introduction}
Logic synthesis optimizes logic networks under various metrics (e.g., area, power, and delay). It is an integral part of modern \emph{Electronic Design Automation}~(EDA) flows.
Combinational logic synthesis optimizes logic networks while preserving the combinational equivalence.  
Even if a logic network has sequential elements, combinational logic synthesis can operate on them by considering register inputs/outputs as primary outputs/inputs, while disregarding constraints on the reachable states.
Sequential logic synthesis aims instead at optimizing logic networks with sequential elements and can be seen as a stronger type of logic synthesis, since it can take into account the fact that not all combinations of register values are reachable in general.
It is widely known that sequential logic synthesis is able to explore a larger solution space and generally provides better \emph{power, performance, and area}~(PPA)~\cite{scorr}. Such PPA opportunities become even more important to grasp as today's chip design cost continues to increase~\cite{tsmc}. 

Previous approaches have been considered for sequential logic synthesis~\cite{scorr,singh1992sis,leiserson1991retiming, savoj2015m, case2011optimal, merge, iwls07, kravets2009sequential,de1991synchronous}.
One such approach is to integrate combinational optimizations together with retiming (i.e., moving registers over combinational nodes)~\cite{leiserson1991retiming,hurst2007fast}, which can exploit optimization opportunities arising due to structural properties across register boundaries~\cite{iwls07}.  
A powerful yet scalable state-of-the-art approach is given by the \emph{sequential SAT-sweeping}~(SSW) algorithm from~\cite{scorr}. This method makes use of SAT and sequential induction~\cite{corr1,corr2, corr3} to prove the validity of logic transformations (i.e., merging sequentially~equivalent~nodes).

In this work, we present a novel scalable algorithm for don't cares based sequential logic synthesis. Our method is based on $k$-step induction, and is orthogonal to the one in~\cite{scorr}. As a matter of fact, our new method applies both redundancy removal and resubstitution while using \emph{sequential observability don't cares}~(SODCs). 
In combinational logic synthesis, \emph{observability don't cares}~(ODCs), i.e., the input patterns for which the value of a wire is not observed at outputs, can be used to find better optimization opportunities ~\cite{MBJJ11, MB06,testa2020extending,lee2021simulation}.

The use of ODCs for optimizations comes with inherent challenges as ODCs can change after applying an ODC-based optimization.
In sequential optimizations, the challenges are more prominent as it is highly non-trivial to consider the sequential nature of circuits while dealing with ODC dependencies.
Although the dependency issues can be avoided by considering only \emph{compatible ODCs}~(CODCs)~\cite{saluja2004robust}, i.e., ODCs that can be used independently at each node, it can lead to many missed optimization opportunities.
Nevertheless, the method we propose utilizes the full power of ODCs in sequential optimizations.
This is achieved by using an inductive approach that takes the reachable states into account and performs simultaneous, in-place optimization of two networks using an ODC-based, combinational optimization method built on Boolean Satisfiability (SAT)~\cite{marques2000boolean}, which uses windowing for scalability.

Essentially, our method naturally finds valid ODC-based sequential optimizations that are compatible with each other, without limiting the search-space to CODC-based optimizations. Thus, it may find better optimizations especially in circuits with sequential feedback that can be problematic for traditional retiming based methods. 
In essence, our method finds sequential optimizations that were never covered by prior approaches in a scalable manner.
Moreover, as our approach does not move registers over combinational logic, the verification of the optimized networks is more likely to succeed, and hence our approach is more attractive for deployment in industrial tools.

The proposed method has been implemented in an industrial tool and shown to achieve 6.9\% average reduction in area after technology mapping on top of state-of-the-art sequential optimization methods (e.g., SSW). All designs were verified using state-of-the-art sequential verification tools.

The organization of the paper: \Cref{sec:background} discusses some background and relevant prior work. \Cref{sec:method} presents a motivating example for sequential synthesis with ODCs followed by our novel scalable sequential logic synthesis approach.
Finally, \cref{sec:results} presents experimental results and \cref{sec:conclusion} concludes with a brief discussion of the results and future work.

\section{Background}
\label{sec:background}
In this section, we provide some background that will be useful to better understand the rest of the paper and briefly describe some state-of-the-art prior work in sequential synthesis.

\subsection{Boolean Network}
A Boolean network is a \emph{directed acyclic graph}~(DAG) representation of a logic network where the nodes correspond to logic gates and edges represent the connections between gates.
A node function can be arbitrary, and usually encoded using its \emph{sum-of-products}~(SOPs) representation or its truth table with respect to node inputs.
The source nodes of the DAG correspond to \emph{primary inputs}~(PIs) or \emph{register outputs}~(ROs) while the sinks correspond to \emph{primary outputs}~(POs) or \emph{register inputs}~(RIs).
The corresponding RI/RO pairs are usually stored in a separate data-structure together with the respective initial values of the registers.
The fanins (fanouts) of a node \(n\) refers to the set of nodes that drives (driven by) \(n\).
I.e., the fanins of \(n\) have directed edges from them to \(n\) and the fanouts have directed edges from \(n\) to them.
The \emph{transitive fanin}~(TFI) cone of a node \(n\) is the set of all nodes from which \(n\) is reachable via a directed path.
Similarly, the \emph{transitive fanout}~(TFO) cone is the set of all nodes reachable from \(n\) via a directed path.

\subsection{Sequential Redundancy}
In logic synthesis, a redundancy is a node or a wire whose value is stuck at a constant in all observable input (PI/RO) patterns. 
A redundant wire can be optimized away by removing the origin node of the wire from the fanin set of the destination node and modifying the destination node's function accordingly.
A node is redundant if all its outgoing wires are redundant.
A sequential redundancy is a generalization of a redundancy where the stuck-at-constant property holds considering all \emph{observable input patterns} and \emph{reachable states}.

\subsection{Sequential Resubstitution}

In logic synthesis, resubstitution refers to replacing a node \(n\) with a different node \(m \neq n\).
In general, \(m\) can be any existing node that is not in the TFO cone of \(n\), or it can be a new node constructed by combining several other non-TFO nodes (called divisors).
The Boolean resubstitution refers to equivalence preserving resubstitutions that are computed considering Boolean properties such as \emph{don't cares}~(DCs).
Additionally, when the implicit restrictions on reachable states of a sequential logic network are considered during resubstitution, we refer to it as sequential resubstitution.

\subsection{Prior Work on Sequential Synthesis} 

A common optimization approach in early works on sequential synthesis is to use retiming together with logic transformations~\cite{de1991synchronous,iwls07}.
In this approach, first the registers are moved around, then the resulting circuit is optimized using combinational methods, and finally retiming is performed again to minimize the register count.
During retiming, if some reconvergent paths have varying numbers of registers, the usual practice is to remove such paths by duplicating the shared nodes, considering small blocks of logic.
In contrast, our SODC-based approach does not move registers, thus it avoids the duplication requirement of shared logic.
Moreover, our approach scales well to much larger logic blocks.

Another prominent sequential optimization method is \emph{sequential SAT-sweeping (SSW)}, which is a generalization of combinational SAT-sweeping~\cite{MCJB05} to the sequential setting, where the idea is to merge sequentially equivalent nodes.
If two nodes \(m\) and \(n\) are equivalent under all observable input patterns of \(n\) in all reachable states, \(n\) can be merged with \(m\) by transferring the fanouts of \(n\) to \(m\), without changing the overall output function of the network.
An efficient SSW algorithm is proposed in~\cite{scorr} where the sequential equivalences among nodes are proven using \emph{bounded model checking}~(BMC)~\cite{biere1999symbolic} and SAT together with induction~\cite{corr1,corr2,corr3}.
Once the equivalence classes are identified, all nodes in a class are merged to a chosen representative node and the dangling nodes are removed.
Despite its practical success, SSW misses many optimizations made possible due to SODCs.
Notably, SSW cannot optimize the simple sequential logic network in \cref{fig:seq}(a) into the one in \cref{fig:seq}(b).

Additionally, Case et al.~\cite{merge} considered a simulation based approach to find merge candidates considering ODCs.
Namely, the network is simulated with random bit patterns to identify node pairs \(a, b\) such that for each simulated pattern, either \(a\) and \(b\) are equal or all paths from \(b\) to combinational outputs are non-controlling. 
Then a new network is created with all candidates merged, the equivalence of the new and original network is proven/disproven using SAT, and if disproved, the  merge candidates are refined.
However, this approach does not scale well to large networks due to large miters used in equivalence checking and hence misses many optimization opportunities.
In contrast, we use a window based approach and check the validity of each optimization in isolation; hence the SAT-based validity checks are scalable.

The method we propose in the next section is able to find optimizations that were never found by the prior approaches.

\section{Scalable Sequential Optimization}
\label{sec:method}
In this section, we first give a brief motivation to our proposed method and introduce sequential induction. Then, we discuss our novel sequential optimization approach in detail.

\subsection{Motivation}
\label{sec:motivation}

\begin{figure}[h]
\centering
\begin{tikzpicture} [every circuit symbol/.style={scale=0.65}, transform shape, scale=0.5]
\Large

\def\nodesep{0.3}

\ctikzset{logic ports/scale=0.7}
\ctikzset{logic ports/thickness=1.1}
\ctikzset{logic ports=ieee,ieeestd ports/pin length=0.75}

\node[and port, anchor=in 1](AND1) at (0, 0) {};

\node at (AND1.bin 2)[ocirc,scale=1.5, left]{};
\draw (AND1.in 1) node[ anchor=east](IN1){$a$};
\draw (AND1.in 2) node[ anchor=east](IN2){$b$};

\draw (AND1.out) -- ++(\nodesep,0) node[and port, anchor=in 1](AND3){} node[midway, above] {$g_1$};
\draw (AND3.in 2) node[ anchor=east](IN3){$d$};
\draw (AND3.out) -- ++(\nodesep,0) node[or port, anchor=in 1](OR1){} node[midway, above]{$w_1$};
\draw (OR1.in 2) node[ anchor=east](IN4){$e$};
draw (AND3.out) -- ++(\nodesep,0) node[or port, anchor=in 1](OR1){};
\draw (OR1.out) -- ++(0.1,0) node[and port, anchor=in 1](AND4){} node[midway, above]{$w_2$};
\draw (AND4.out) node[ anchor=west](O1){$o_1$};

\node[and port, anchor=in 1](AND2) at (0, -1.4) {};

 \node at (AND2.bin 1)[ocirc, scale=1.5, left]{};
 \draw (AND2.in 1) node[ anchor=east](IN5){ $a$};
 \draw (AND2.in 2) node[ anchor=east](IN6){ $b$};
 \draw (AND2.out) node[above, anchor = south west ] { $g_2$} -- ++(\nodesep * 12, 0) node[ anchor=base](TEMP1){} |- (AND4.in 2) ;

\node[and port, anchor=in 1](AND5) at (9, -0.7) {};

 \node at (AND5.bin 1)[ocirc, scale=1.5, left]{};
 \draw (AND5.in 1) node[ anchor=east](IN7){$a$};
 \draw (AND5.in 2) node[ anchor=east](IN8){$b$};
 \draw (AND5.out) -- ++(\nodesep,0) node[and port, anchor=in 1](AND6){} node[midway, above] {$g_2$};
 \draw (AND6.in 2) node[ anchor=east](IN9){$e$};
 \draw (AND6.out) node[ anchor=west](O2){$o_1$};

\node (a) at (3, -2.5) [draw,draw=none] {$\textbf{(a)}$};
\node (b) at (10.5, -2.5) [draw,draw=none] {$\textbf{(b)}$};

\end{tikzpicture}
\caption{ A combinational logic network (a) and its optimized version~(b). }
\label{fig:comb}
\end{figure}
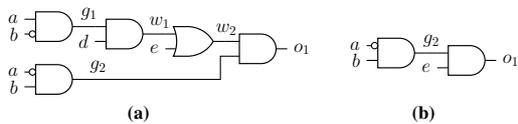

Consider the purely combinational logic network shown in~\cref{fig:comb}(a) and observe that the wires \(g_1\) and \(g_2\) can never be 1 at the same time.
This implies that whenever \(g_2 = 1\), \(g_1\) must be \(0\).
Since \(w_2\) is observed at output \(o_1\) only when \(g_2 = 1\), one can simplify the circuit by assuming that \(g_1\) is stuck at \(0\), which, in turn, implies that \(w_1\) is also stuck at \(0\).
This leads to the optimized circuit in \cref{fig:comb}(b).

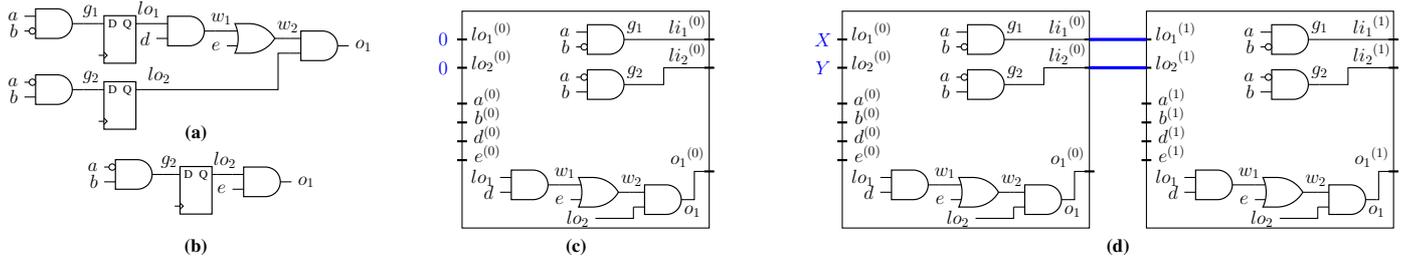
\begin{figure*}[t]
\centering
\begin{tikzpicture} [every circuit symbol/.style={scale=0.65}, transform shape, scale=0.5]

\Large

\def\nodesep{0.3}
\def\ffsep{0.4}
\def\pinoffset{0.3}

\tikzset{ffD/.style={flipflop, flipflop def={t1=D, t6=Q, c3 = 1}}}

\ctikzset{flipflops/scale=0.5}
\ctikzset{logic ports/scale=0.7}
\ctikzset{flipflops/thickness=1.1}
\ctikzset{logic ports/thickness=1.1}
\ctikzset{logic ports=ieee,ieeestd ports/pin length=0.75}

\newcommand{\seqframe}[3]{
    \begin{scope}[xshift=#2cm,yshift=#3cm]
 \draw (0, 0) rectangle (6.5, 5.75);
 \node(f#1-lo1-ext) at ( -\pinoffset, 5) {};
 \node(f#1-lo1-int) at (  \pinoffset, 5) {};
 \draw [thick](f#1-lo1-ext) -- (f#1-lo1-int);
 \node() at ($(f#1-lo1-int.north east)+ (0.3, 0)$) {${lo_1}^{(#1)}$};
 
 \node(f#1-lo2-ext) at ( -\pinoffset, 4.25) {};
 \node(f#1-lo2-int) at (  \pinoffset, 4.25) {};
 \draw [thick](f#1-lo2-ext) -- (f#1-lo2-int);
 \node() at ($(f#1-lo2-int.north east) + (0.3, 0)$) {${lo_2}^{(#1)}$};

 \node(f#1-a-ext) at ( -\pinoffset, 3.3) {};
 \node(f#1-a-int) at (  \pinoffset, 3.3) {};
 \draw [thick](f#1-a-ext) -- (f#1-a-int);
 \node() at ($(f#1-a-int.north east)+ (0.2, 0)$) {$a^{(#1)}$};
 
 \node(f#1-b-ext) at ( -\pinoffset, 2.8) {};
 \node(f#1-b-int) at (  \pinoffset, 2.8) {};
 \draw [thick](f#1-b-ext) -- (f#1-b-int);
 \node() at ($(f#1-b-int.north east)+ (0.2, 0)$) {$b^{(#1)}$};

 \node(f#1-d-ext) at ( -\pinoffset, 2.3) {};
 \node(f#1-d-int) at (  \pinoffset, 2.3) {};
 \draw [thick](f#1-d-ext) -- (f#1-d-int);
 \node() at ($(f#1-d-int.north east)+ (0.2, 0)$) {$d^{(#1)}$};

 \node(f#1-e-ext) at ( -\pinoffset, 1.8) {};
 \node(f#1-e-int) at (  \pinoffset, 1.8) {};
 \draw [thick](f#1-e-ext) -- (f#1-e-int);
 \node() at ($(f#1-e-int.north east)+ (0.2, 0)$) {$e^{(#1)}$};
 
 \node(f#1-li1-ext) at ( 6.5 +\pinoffset, 5) {};
 \node(f#1-li1-int) at ( 6.5 -\pinoffset, 5) {};
 \draw [thick](f#1-li1-ext) -- (f#1-li1-int);
 \node() at ($(f#1-li1-int.north west)+ (-0.1, 0.2)$) {${li_1}^{(#1)}$};

 \node(f#1-li2-ext) at ( 6.5 +\pinoffset, 4.25) {};
 \node(f#1-li2-int) at ( 6.5 -\pinoffset, 4.25) {};
 \draw [thick](f#1-li2-ext) -- (f#1-li2-int);
 \node() at ($(f#1-li2-int.north west)+ (-0.1, 0.2)$) {${li_2}^{(#1)}$};

 \node(f#1-po1-ext) at ( 6.5 +\pinoffset, 1.5) {};
 \node(f#1-po1-int) at ( 6.5 -\pinoffset, 1.5) {};
 \draw [thick](f#1-po1-ext) -- (f#1-po1-int);
 \node() at ($(f#1-po1-int.north west)+ (-0.1, 0.2)$) {${o_1}^{(#1)}$};

 \node[and port, anchor=in 1](f#1-n1) at (3, 5.195) {};
 \node at (f#1-n1.bin 2)[ocirc,scale=1.5, left]{};
 \draw (f#1-n1.in 1) node[ anchor=east](IN1){$a$};
 \draw (f#1-n1.in 2) node[ anchor=east](IN2){$b$};
  \draw (f#1-n1.out)node[above] {$g_1$} -- ++(0.5, 0) -- (f#1-li1-ext) ;

 \node[and port, anchor=in 1](f#1-n2) at (3, 4) {};
 \node at (f#1-n2.bin 1)[ocirc,scale=1.5, left]{};
 \draw (f#1-n2.in 1) node[ anchor=east](IN1){$a$};
 \draw (f#1-n2.in 2) node[ anchor=east](IN2){$b$};
 \draw (f#1-n2.out)node[above] {$g_2$} -- ++(0.75, 0) |- (f#1-li2-ext) ;

\node[and port, anchor=in 1](f#1-n3) at (1, 1.34) {};
\draw (f#1-n3.in 1) node[ anchor=east](IN1){$lo_1$};
\draw (f#1-n3.in 2) node[ anchor=east](IN2){$d$};

 \draw (f#1-n3.out) -- ++(\nodesep,0) node[or port, anchor=in 1](f#1-n4){} node[midway, above]{$w_1$};
 \draw (f#1-n4.out) -- ++(0.1,0) node[and port, anchor=in 1](f#1-n5){}  node[midway, above]{$w_2$};

\draw (f#1-n4.in 2) node[ anchor=east](i20){$e$};
\draw (f#1-n5.in 2)  -- ++(0,-0.3) |- ++ (-1,0) node[ anchor=east](i21){$lo_2$};
\draw (f#1-n5.out)node[below](O21){$o_1$} -| ++ (0.05, 0)|- (f#1-po1-ext);

 \end{scope}
}

\draw (0,0) node[ffD](D1){};
\draw (D1.pin 1) -- ++(-\ffsep,0) node[and port, anchor=out](AND1){} node[midway, above]{$g_1$};
\node at (AND1.bin 2)[ocirc,scale=1.5, left]{};
\draw (AND1.in 1) node[ anchor=east](IN1){$a$};
\draw (AND1.in 2) node[ anchor=east](IN2){$b$};

\draw (D1.pin 6) -- ++(\ffsep,0) node[and port, anchor=in 1](AND3){} node[midway,above] {${lo}_1$};
\draw (AND3.in 2) node[ anchor=east](IN3){$d$};
\draw (AND3.out) -- ++(\nodesep,0) node[or port, anchor=in 1](OR1){} node[midway, above]{$w_1$};
\draw (OR1.in 2) node[ anchor=east](IN4){$e$};
draw (AND3.out) -- ++(\nodesep,0) node[or port, anchor=in 1](OR1){};
\draw (OR1.out) -- ++(0.1,0) node[and port, anchor=in 1](AND4){} node[midway, above]{$w_2$};
\draw (AND4.out) node[ anchor=west](O1){$o_1$};

 \draw (0,-1.75) node[ffD](D2){};
 \draw (D2.pin 1) -- ++(-\ffsep,0) node[and port, anchor=out](AND2){}node[midway, above]{$g_2$} ;
 \node at (AND2.bin 1)[ocirc, scale=1.5, left]{};
 \draw (AND2.in 1) node[ anchor=east](IN5){ $a$};
 \draw (AND2.in 2) node[ anchor=east](IN6){ $b$};
 \draw (D2.pin 6) node[above, anchor = south west ] { ${lo}_2$} -- ++(\nodesep * 12, 0) node[ anchor=base](TEMP1){} |- (AND4.in 2) ;

 \draw (2,-4) node[ffD](D3){};
 \draw (D3.pin 1) -- ++(-\nodesep,0) node[and port, anchor=out](AND5){} node[midway, above]{$g_2$} ;
 \node at (AND5.bin 1)[ocirc, scale=1.5, left]{};
 \draw (AND5.in 1) node[ anchor=east](IN7){$a$};
 \draw (AND5.in 2) node[ anchor=east](IN8){$b$};
 \draw (D3.pin 6) -- ++(\ffsep,0) node[and port, anchor=in 1](AND6){} node[midway, above] {${lo}_2$};
 \draw (AND6.in 2) node[ anchor=east](IN9){$e$};
 \draw (AND6.out) node[ anchor=west](O2){$o_1$};

\seqframe{0}{9}{-5};
\seqframe{0}{19}{-5};
\seqframe{1}{27}{-5};

\draw(25.5,0)[color=blue, very thick] to (27,0);
\draw(25.5,-0.75)[color=blue, very thick] -- (27,-0.75);

\node (x)[color=blue] at (8.5,0) {$0$} ;
\node (y)[color=blue] at (8.5,-0.75) {$0$} ;

\node (x)[color=blue] at (18.5,0) {$X$} ;
\node (y)[color=blue] at (18.5,-0.75) {$Y$} ;

\node (a) at (2, -2.5) [draw,draw=none] {$\textbf{(a)}$};
\node (b) at (2, -5.5) [draw,draw=none] {$\textbf{(b)}$};
\node (c) at (12, -5.5) [draw,draw=none] {$\textbf{(c)}$};
\node (d) at (26.25, -5.5) [draw,draw=none] {$\textbf{(d)}$};
\end{tikzpicture}
\caption{A sequential logic network (a), its optimized version (b), and its base case network (c) and the inductive case network (d) for 1-step sequential induction.}
\label{fig:seq}
\end{figure*}

Now consider the sequential circuit in \cref{fig:seq}(a) which is same as the one in \cref{fig:comb}(a) except for the two registers at \(g_1\) and \(g_2\).
If we consider this as a combinational network (i.e., disregard the registers, consider \(g_1,g_2\) to be POs, and consider \(lo_1,lo_2\) to be PIs), the previous reasoning no longer applies; \(lo_1, lo_2\) can take arbitrary values, and hence \(lo_1\) is observed even when it is 1.
However, if we additionally know that the initial values of \(lo_1, lo_2\) are \((0, 0)\) (or any combination of values different from \((1,1)\)), the optimization is still possible, as by design, \(lo_1\) and  \(lo_2\) can never be \(1\) at the same time in the subsequent clock cycles. The optimized circuit is shown in \cref{fig:seq}(b). The state-of-the-art sequential optimization routines such as \emph{scorr}, \emph{lcorr}, \emph{scl}, and \emph{retime} of the logic synthesis tool ABC~\cite{abc-paper} cannot find this optimization.

The goal of the proposed method is to identify this kind of optimization opportunities in sequential logic networks in a scalable way. 
We remark that, while a retiming based optimization method might be able to optimize the example above, such methods perform poorly especially when there is sequential feedback (e.g., finite state machines) or varying numbers of registers along different reconvergent logic paths.

The optimization in the example holds due to two facts:
\begin{enumerate}
    \item (\emph{Reachability}) Not all states (value combinations for the sequential elements) are reachable.
    \item (\emph{Observability}) In all reachable states, the optimization is valid due to the ODCs.
\end{enumerate}
These two facts together form a notion of \emph{sequential ODCs}~(SODCs), a generalization of ODCs in combinational logic networks into the sequential setting.

\subsection{Framework Definition}
\label{sec:framework-def}
To use SODCs in optimization, we first take the \emph{reachability} of states into account. To this end, a widely used technique is to use the so-called sequential induction~\cite{corr1,corr2} which considers two combinational networks that are derived from the original network according to Definition~\ref{def:derived}. 

\newtheorem{define}{Definition}
\begin{define}
\label{def:derived}
For a network N, the \(k\)-step base case network \(N^b\) is the combinational network obtained taking \(k\) copies of \(N\), which are referred to as frames, connecting RIs of each frame to the corresponding ROs of the subsequent frame, replacing the ROs of first frame with the respective register initial values, and designating RIs of the last frame as POs.
The \(k\)-step inductive case network \(N^i\) is similarly defined except we consider \(k+1\) frames and consider ROs of the first frame as PIs.
\end{define}

For the example network of \cref{fig:seq}(a), the base case and the inductive case networks for 1-step sequential induction are shown in \cref{fig:seq}(c) and \cref{fig:seq}(d) respectively.
The behavior of \(N^b\) is the same as that of the original network \(N\) for the initial clock cycle
and the behavior of \(N^i\) is the same as two consecutive clock cycles of \(N\) for any initial state.
If an optimization is valid in all frames of the base case and the last frame of the inductive case, then it is a valid sequential optimization for the original network (see \cref{thm:induction}).

On top of the reachability criterion, we consider the \emph{observability} to identify sequential optimization opportunities.
It seems straightforward to consider the sets of ODC based optimizations in the base case and inductive networks and then take the intersection of the two sets as the final set of optimizations. However, as discussed in \cref{sec:introduction}, this approach only works with CODCs which does not have dependencies among them.
Unfortunately, using CODCs in place of ODCs lead to many missed optimization opportunities.
The regular ODCs can have dependencies in them, and cause this simple algorithm to fail.
In the remainder of this section, we present an algorithm that, by design, avoids dependency issues of regular ODCs without falling back to CODCs.

\subsection{Proposed Method}
Our proposed algorithm is based on sequential induction and it can fully utilize ODCs by simultaneously optimizing base case and inductive case networks.

Namely, we start by constructing the base case and inductive case networks for sequential induction.
Then, considering one node at a time, we check if there is a valid optimization for that node in both the base case and the inductive case networks.
If so, we immediately update both the derived networks as well as the original network.
This approach allows the algorithm to find subsequent optimizations for the remaining nodes that may depend on the already applied optimizations,
and it avoids any dependency issues that would arise if we were to use the simple approach we stated at the end of \cref{sec:framework-def} with regular ODCs. 
Hence, the algorithm computes compatible sequential optimizations without limiting to CODCs.

The algorithm considers fanin redundancies for each gate \(n\) in the network.
Namely, for each fanin \(f\) of \(n\) we check whether \(f\) is effectively stuck at 0 or 1 in the base case network and the last frame of the inductive case network.
Since both the base case and inductive case networks are purely combinational, it is possible to use any
combinational redundancy check for this purpose.
To this end, let \(n'\) be the node obtained by fixing fanin \(f\) of \(n\) at the target constant value.
In our implementation, we check if replacing \(n\) with \(n'\) is valid  using a SAT problem.
If the problem is unsatisfiable~(UNSAT), then the optimization is valid.
To make the overall algorithm scalable, we optimize the SAT formulation not to consider all POs and RIs,
but instead consider the leaf nodes of a small TFO cone rooted at \(n\).
If the optimization is shown to be valid for both the base and inductive case networks, then we apply it in both the networks as well as in the original network (see \cref{sec:algorithm} for details).

As an illustrative example, consider \cref{fig:seq}. 
We can prove, for example, that \(w_1\) is stuck at 0, in both the base case and the inductive case networks, which will result in the optimized network in \cref{fig:seq}(b) (assuming all registers are initially 0).

We consider three enhancement on our proposed method.

\subsubsection{Enhancement 1} We extend our algorithm to use \(k\)-step sequential induction where the base case network has \(k\) frames and the inductive case network has \(k+1\) frames.
In this case, we check if the target redundancy is valid in all \(k\) frames of the base case and the last frame of the inductive case.
The validity in all frames of the base case implies that the original network would behave the same for first \(k\) clock cycles (starting with the initial state).
The validity in the last frame of the inductive case implies that the original network will behave the same for any \(k+1\) clock cycles, starting from any state.

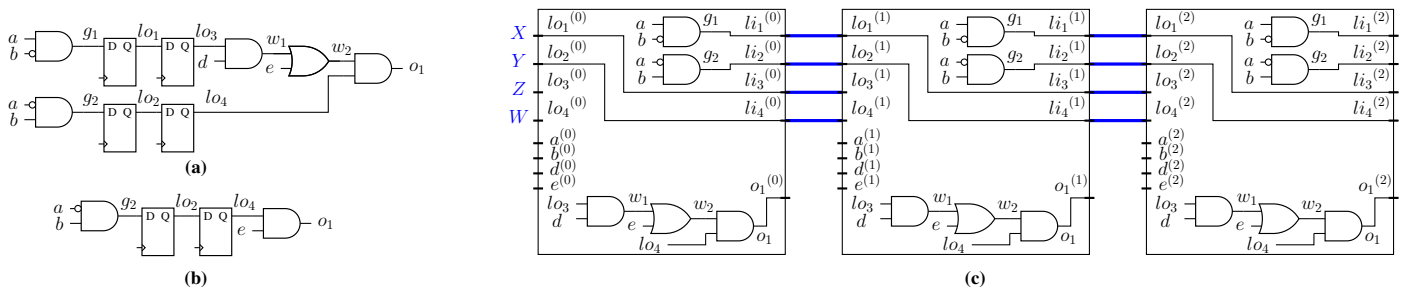
\begin{figure*}[b]
\centering
\begin{tikzpicture} [every circuit symbol/.style={scale=0.65}, transform shape, scale=0.5]
\Large

\def\nodesep{0.2}
\def\ffsep{0.4}
\def\pinoffset{0.3}

\tikzset{ffD/.style={flipflop, flipflop def={t1=D, t6=Q, c3 = 1}}}

\ctikzset{flipflops/scale=0.5}
\ctikzset{logic ports/scale=0.7}
\ctikzset{flipflops/thickness=1.1}
\ctikzset{logic ports/thickness=1.1}
\ctikzset{logic ports=ieee,ieeestd ports/pin length=0.75}
\newcommand{\seqframe}[3]{
    \begin{scope}[xshift=#2cm,yshift=#3cm]
 \draw (0, 0) rectangle (6.5, 6.5);
 \node(f#1-lo1-ext) at ( -\pinoffset, 5.8) {};
 \node(f#1-lo1-int) at (  \pinoffset, 5.8) {};
 \draw [thick](f#1-lo1-ext) -- (f#1-lo1-int);
 \node() at ($(f#1-lo1-int.north east)+ (0.3,0.2) $) {${lo_1}^{(#1)}$};
 
 \node(f#1-lo2-ext) at ( -\pinoffset, 5.05) {};
 \node(f#1-lo2-int) at (  \pinoffset, 5.05) {};
 \draw [thick](f#1-lo2-ext) -- (f#1-lo2-int);
 \node() at ($(f#1-lo2-int.north east)+ (0.3,0.2) $) {${lo_2}^{(#1)}$};

 \node(f#1-lo3-ext) at ( -\pinoffset, 4.3) {};
 \node(f#1-lo3-int) at (  \pinoffset, 4.3) {};
 \draw [thick](f#1-lo3-ext) -- (f#1-lo3-int);
 \node() at ($(f#1-lo3-int.north east)+ (0.3,0.2) $) {${lo_3}^{(#1)}$};

 \node(f#1-lo4-ext) at ( -\pinoffset, 3.55) {};
 \node(f#1-lo4-int) at (  \pinoffset, 3.55) {};
 \draw [thick](f#1-lo4-ext) -- (f#1-lo4-int);
 \node() at ($(f#1-lo4-int.north east)+ (0.3,0.2) $) {${lo_4}^{(#1)}$};

 \node(f#1-a-ext) at ( -\pinoffset, 2.95) {};
 \node(f#1-a-int) at (  \pinoffset, 2.95) {};
 \draw [thick](f#1-a-ext) -- (f#1-a-int);
 \node() at ($(f#1-a-int.north east)+ (0.2,0) $) {$a^{(#1)}$};
 
 \node(f#1-b-ext) at ( -\pinoffset, 2.55) {};
 \node(f#1-b-int) at (  \pinoffset, 2.55) {};
 \draw [thick](f#1-b-ext) -- (f#1-b-int);
 \node() at ($(f#1-b-int.north east)+ (0.2,0) $) {$b^{(#1)}$};

 \node(f#1-d-ext) at ( -\pinoffset, 2.15) {};
 \node(f#1-d-int) at (  \pinoffset, 2.15) {};
 \draw [thick](f#1-d-ext) -- (f#1-d-int);
 \node() at ($(f#1-d-int.north east)+ (0.2,0) $) {$d^{(#1)}$};

 \node(f#1-e-ext) at ( -\pinoffset, 1.75) {};
 \node(f#1-e-int) at (  \pinoffset, 1.75) {};
 \draw [thick](f#1-e-ext) -- (f#1-e-int);
 \node() at ($(f#1-e-int.north east) + (0.2,0) $) {$e^{(#1)}$};
 
 \node(f#1-li1-ext) at ( 6.5 +\pinoffset, 5.8) {};
 \node(f#1-li1-int) at ( 6.5 -\pinoffset, 5.8) {};
 \draw [thick](f#1-li1-ext) -- (f#1-li1-int);
 \node() at ($(f#1-li1-int.north west) + (-0.1,0.2) $) {${li_1}^{(#1)}$};

 \node(f#1-li2-ext) at ( 6.5 +\pinoffset, 5.05) {};
 \node(f#1-li2-int) at ( 6.5 -\pinoffset, 5.05) {};
 \draw [thick](f#1-li2-ext) -- (f#1-li2-int);
 \node() at ($(f#1-li2-int.north west) + (-0.1, 0.2)$) {${li_2}^{(#1)}$};

 \node(f#1-li3-ext) at ( 6.5 +\pinoffset, 4.3) {};
 \node(f#1-li3-int) at ( 6.5 -\pinoffset, 4.3) {};
 \draw [thick](f#1-li3-ext) -- (f#1-li3-int);
 \node() at ($(f#1-li3-int.north west) + (-0.1,0.2) $) {${li_3}^{(#1)}$};

 \node(f#1-li4-ext) at ( 6.5 +\pinoffset, 3.55) {};
 \node(f#1-li4-int) at ( 6.5 -\pinoffset, 3.55) {};
 \draw [thick](f#1-li4-ext) -- (f#1-li4-int);
 \node() at ($(f#1-li4-int.north west) + (-0.1, 0.2)$) {${li_4}^{(#1)}$};

 \node(f#1-po1-ext) at ( 6.5 +\pinoffset, 1.5) {};
 \node(f#1-po1-int) at ( 6.5 -\pinoffset, 1.5) {};
 \draw [thick] (f#1-po1-ext) -- (f#1-po1-int);
 \node() at ($(f#1-po1-int.north west) + (0, 0.2)$) {${o_1}^{(#1)}$};

 \node[and port, anchor=in 1](f#1-n1) at (3, 6.1) {};
 \node at (f#1-n1.bin 2)[ocirc,scale=1.5, left]{};
 \draw (f#1-n1.in 1) node[ anchor=east](IN1){$a$};
 \draw (f#1-n1.in 2) node[ anchor=east](IN2){$b$};
  \draw (f#1-n1.out)node[above] {$g_1$} -- ++(0.5, 0) |- (f#1-li1-ext) ;

 \node[and port, anchor=in 1](f#1-n2) at (3, 5.1) {};
 \node at (f#1-n2.bin 1)[ocirc,scale=1.5, left]{};
 \draw (f#1-n2.in 1) node[ anchor=east](IN1){$a$};
 \draw (f#1-n2.in 2) node[ anchor=east](IN2){$b$};
 \draw (f#1-n2.out)node[above] {$g_2$} -- ++(0.5, 0) |- (f#1-li2-ext) ;

\node[and port, anchor=in 1](f#1-n3) at (1, 1.34) {};
\draw (f#1-n3.in 1) node[ anchor=east](IN1){$lo_3$};
\node (d) at (0.5,0.9) {$d$};

 \draw (f#1-n3.out) -- ++(\nodesep,0) node[or port, anchor=in 1](f#1-n4){} node[midway, above]{$w_1$};
 \draw (f#1-n4.out) -- ++(0.1,0) node[and port, anchor=in 1](f#1-n5){}  node[midway, above]{$w_2$};

\draw (f#1-n4.in 2) node[ anchor=east](i20){$e$};
\draw (f#1-n5.in 2)  -- ++(0,-0.3) |- ++ (-1,0) node[ anchor=east](i21){$lo_4$};
\draw (f#1-n5.out) node[below](O21){$o_1$}-| ++ (0.05, 0)|- (f#1-po1-ext);
\draw (f#1-lo1-ext) -- (2.25,5.8) |- (f#1-li3-ext);
\draw (f#1-lo2-ext) -- (1.75,5.05) |- (f#1-li4-ext);

 \end{scope}
}

\draw (0,0) node[ffD](D0){};

\draw (D0.pin 1) -- ++(-\ffsep,0) node[and port, anchor=out](AND1){} node[midway, above]{$g_1$};
\node at (AND1.bin 2)[ocirc,scale=1.5, left]{};
\draw (AND1.in 1) node[ anchor=east](IN1){$a$};
\draw (AND1.in 2) node[ anchor=east](IN2){$b$};

\draw (D0.pin 6) -- ++(\ffsep,0) node[ffD, anchor=pin 1](D1){} node[midway,above] {${lo}_1$};

\draw (D1.pin 6) -- ++(\ffsep,0) node[and port, anchor=in 1](AND3){} node[midway,above] {${lo}_3$};
\draw (AND3.in 2) node[ anchor=east](IN3){$d$};
\draw (AND3.out) -- ++(\nodesep,0) node[or port, anchor=in 1](OR1){} node[midway, above]{$w_1$};
\draw (OR1.in 2) node[ anchor=east](IN4){$e$};
\draw (AND3.out) -- ++(\nodesep,0) node[or port, anchor=in 1](OR1){};
\draw (OR1.out) -- ++(0.1,0) node[and port, anchor=in 1](AND4){}node[midway, above]{$w_2$};
\draw (AND4.out) node[ anchor=west](O1){$o_1$};

 \draw (0,-1.75) node[ffD](D02){};
 \draw (D02.pin 1) -- ++(-\ffsep,0) node[and port, anchor=out](AND2){}node[midway, above]{$g_2$} ;
 \node at (AND2.bin 1)[ocirc, scale=1.5, left]{};
 \draw (AND2.in 1) node[ anchor=east](IN5){ $a$};
 \draw (AND2.in 2) node[ anchor=east](IN6){ $b$};

 \draw (D02.pin 6) -- ++(\ffsep,0) node[ffD, anchor=pin 1](D2){} node[midway,above] {${lo}_2$};
 
 \draw (D2.pin 6) node[above, anchor = south west ] { ${lo}_4$} -- ++(\nodesep * 17, 0) node[ anchor=base](TEMP1){} |- (AND4.in 2) ;

 \draw (1,-4.5) node[ffD](D03){};
 
 \draw (D03.pin 1) -- ++(-\nodesep,0) node[and port, anchor=out](AND5){}node[above]{$g_2$} ;
 \node at (AND5.bin 1)[ocirc, scale=1.5, left]{};
 
 \draw (AND5.in 1) node[ anchor=east](IN7){$a$};
 \draw (AND5.in 2) node[ anchor=east](IN8){$b$};

  \draw (D03.pin 6) -- ++(\ffsep,0) node[ffD, anchor=pin 1](D3){} node[midway,above] {${lo}_2$};
 \draw (D3.pin 6) -- ++(\ffsep,0) node[and port, anchor=in 1](AND6){} node[midway, above] {${lo}_4$};

 \draw (AND6.in 2) node[ anchor=east](IN9){$e$};
 \draw (AND6.out) node[ anchor=west](O2){$o_1$};

\draw(17.5,0.7)[color=blue, very thick] to (19,0.7);
\draw(17.5,-0.05)[color=blue, very thick] -- (19,-0.05);
\draw(17.5,-0.8)[color=blue, very thick] to (19,-0.8);
\draw(17.5,-1.55)[color=blue, very thick] -- (19,-1.55);

\draw(25.5,0.7)[color=blue, very thick] -- (27,0.7);
\draw(25.5,-0.05)[color=blue, very thick] -- (27,-0.05);
\draw(25.5,-0.8)[color=blue, very thick] -- (27,-0.8);
\draw(25.5,-1.55)[color=blue, very thick] -- (27,-1.55);

\node (x)[color=blue] at (10.5,0.8) {$X$} ;
\node (y)[color=blue] at (10.5,0.05) {$Y$} ;
\node (x)[color=blue] at (10.5,-0.7) {$Z$} ;
\node (y)[color=blue] at (10.5,-1.45) {$W$} ;

\seqframe{0}{11}{-5.1};
\seqframe{1}{19}{-5.1};
\seqframe{2}{27}{-5.1};

\node (a) at (2, -2.8) [draw,draw=none] {$\textbf{(a)}$};
\node (b) at (2, -5.75) [draw,draw=none] {$\textbf{(b)}$};
\node (c) at (22.5, -5.75) [draw,draw=none] {$\textbf{(c)}$};

\end{tikzpicture}
\caption{A sequential logic network (a), its optimized version (b), and its inductive case network (c) for 2-step sequential induction.}
\label{fig:seq_k_step}
\end{figure*}

\Cref{fig:seq_k_step}(a) shows an example sequential network which can be optimized to the one in \cref{fig:seq_k_step}(b) with 2-step sequential induction (assuming all registers are initially 0).
In the last frame of the inductive network (\cref{fig:seq_k_step}(c)), \(lo_3, lo_4\) are fed by the gates \(g_1,g_2\) of the first frame, so \(lo_3, lo_4\) of the last frame are never \(1\) at the same time. Thus the algorithm is able to prove that \(w_1\) of the last frame is stuck at zero.
In contrast, if 1-step induction is used, we only get the first two frames of \cref{fig:seq_k_step}(c), and the second frame's \(lo_3, lo_4\) are fed by two arbitrary inputs from the first frame. Hence, all value combinations are possible, so \(w_1\) of the second frame is not stuck at zero.

\subsubsection{Enhancement 2} Our approach is not limited to fanin redundancies but also extends to resubstitutions under ODCs.
Namely, we consider a subset of nodes that are \emph{not} in the TFO cone of the considered node \(n\) as divisors. 
Then we consider as resubstitution candidates all versions of \(n\) obtained by replacing a fanin of \(n\) with a divisor.
As with the redundancies, for each resubstitution candidate \(n'\), we use SAT to check if some window output would differ when \(n\) is replaced with \(n'\).

\subsubsection{Enhancement 3}

We further improve our method by considering redundancy assumptions in the base case and the inductive case networks.
Namely, once a valid optimization \(\Delta\) is found in the first frame of the base case, before checking its validity in \(i\)-th frame, we apply it in all the frames before \(i\).
Once its validity is confirmed in all frames of the base case network, we apply it in the first \(k\) frames of the inductive case network and check for its validity in the last frame.
At any point, if we find \(\Delta\) is not valid for the considered frame, we undo it in all previous frames. 

\begin{figure}[t]
\centering
\begin{tikzpicture} [every circuit symbol/.style={scale=0.65}, transform shape, scale=0.5]
\Large

\def\nodesep{0.2}
\def\ffsep{0.4}
\def\pinoffset{0.3}

\tikzset{ffD/.style={flipflop, flipflop def={t1=D, t6=Q, c3 = 1}}}

\ctikzset{flipflops/scale=0.5}
\ctikzset{logic ports/scale=0.7}
\ctikzset{flipflops/thickness=1.1}
\ctikzset{logic ports/thickness=1.1}
\ctikzset{logic ports=ieee,ieeestd ports/pin length=0.75}

\newcommand{\seqframe}[3]{
    \begin{scope}[xshift=#2cm,yshift=#3cm]
 \draw (0, 0) rectangle (5, 3.5);
 \node(f#1-lo1-ext) at ( -\pinoffset,2.75) {};
 \node(f#1-lo1-int) at (  \pinoffset, 2.75) {};
 \draw [thick](f#1-lo1-ext) -- (f#1-lo1-int);
 \node() at ($(f#1-lo1-int.north east)+(0.3,0)$) {${lo_1}^{(#1)}$};
 
 \node(f#1-a-ext) at ( -\pinoffset, 2) {};
 \node(f#1-a-int) at (  \pinoffset, 2) {};
 \draw [thick](f#1-a-ext) -- (f#1-a-int);
 \node() at ($(f#1-a-int.north)+(0.3,0)$) {$a^{(#1)}$};
 
 \node(f#1-b-ext) at ( -\pinoffset, 1.5) {};
 \node(f#1-b-int) at (  \pinoffset, 1.5) {};
 \draw [thick](f#1-b-ext) -- (f#1-b-int);
 \node() at ($(f#1-b-int.north)+(0.3,0)$) {$b^{(#1)}$};

 \node(f#1-li1-ext) at ( 5 +\pinoffset, 2.75) {};
 \node(f#1-li1-int) at ( 5 -\pinoffset, 2.75) {};
 \draw [thick](f#1-li1-ext) -- (f#1-li1-int);
 \node() at ($(f#1-li1-int.north west)+(-0.01,0.2)$) {${li_1}^{(#1)}$};

 \node(f#1-po1-ext) at ( 5 +\pinoffset, 1.5) {};
 \node(f#1-po1-int) at ( 5 -\pinoffset, 1.5) {};
 \draw [thick](f#1-po1-ext) -- (f#1-po1-int);
 \node() at ($(f#1-po1-int.north west)+(0,0.15)$) {${o_1}^{(#1)}$};

 \node[and port, anchor=in 1](f#1-n1)[color=blue] at (0.9, 0.75) {};
\draw (f#1-n1.out) -- ++(0,0) node[and port, anchor=in 2](f#1-n2){} {[color=blue] node[midway, below]{$g_1$}};
\draw (f#1-n2.out) -- ++(0,0) node[midway, below]{$g_2$};

\draw (f#1-n1.in 1) [color=blue] node[ anchor=east](IN1){$lo_1$};
\draw (f#1-n1.in 2) [color=blue] node[ anchor=east](IN2){$a$};
 \draw (f#1-n2.in 1) node[ anchor=south](IN3){$b$};
\draw (f#1-n2.out) to[short, *-] ++ (0.5, 0)|-(f#1-po1-ext) ;
\draw (f#1-n2.out) -| ++ (0, 0)|- (f#1-li1-ext) ;

 \end{scope}
}

\newcommand{\seqframee}[3]{
    \begin{scope}[xshift=#2cm,yshift=#3cm]
 \draw (0, 0) rectangle (5, 3.5);
 \node(f#1-lo1-ext) at ( -\pinoffset,2.75) {};
 \node(f#1-lo1-int) at (  \pinoffset, 2.75) {};
 \draw [thick](f#1-lo1-ext) -- (f#1-lo1-int);
 \node() at ($(f#1-lo1-int.north east)+(0.3,0)$) {${lo_1}^{(#1)}$};
 
 \node(f#1-a-ext) at ( -\pinoffset, 2) {};
 \node(f#1-a-int) at (  \pinoffset, 2) {};
 \draw [thick](f#1-a-ext) -- (f#1-a-int);
 \node() at ($(f#1-a-int.north)+(0.3,0)$) {$a^{(#1)}$};
 
 \node(f#1-b-ext) at ( -\pinoffset, 1.5) {};
 \node(f#1-b-int) at (  \pinoffset, 1.5) {};
 \draw [thick](f#1-b-ext) -- (f#1-b-int);
 \node() at ($(f#1-b-int.north)+(0.3,0)$) {$b^{(#1)}$};

 \node(f#1-li1-ext) at ( 5 +\pinoffset, 2.75) {};
 \node(f#1-li1-int) at ( 5 -\pinoffset, 2.75) {};
 \draw [thick](f#1-li1-ext) -- (f#1-li1-int);
 \node() at ($(f#1-li1-int.north west)+(-0.01,0.2)$) {${li_1}^{(#1)}$};

 \node(f#1-po1-ext) at ( 5 +\pinoffset, 1.5) {};
 \node(f#1-po1-int) at ( 5 -\pinoffset, 1.5) {};
 \draw [thick](f#1-po1-ext) -- (f#1-po1-int);
 \node() at ($(f#1-po1-int.north west)+(0,0.15)$) {${o_1}^{(#1)}$};

 \node[and port, anchor=in 1](f#1-n1) at (0.9, 0.75) {};
\draw (f#1-n1.out) -- ++(0,0) node[and port, anchor=in 2](f#1-n2){} {node[midway, below]{$g_1$}};
\draw (f#1-n2.out) -- ++(0,0) node[midway, below]{$g_2$};

\draw (f#1-n1.in 1) node[ anchor=east](IN1){$lo_1$};
\draw (f#1-n1.in 2) node[ anchor=east](IN2){$a$};
 \draw (f#1-n2.in 1) node[ anchor=south](IN3){$b$};
\draw (f#1-n2.out) to[short, *-] ++ (0.5, 0)|-(f#1-po1-ext) ;
\draw (f#1-n2.out) -| ++ (0, 0)|- (f#1-li1-ext) ;

 \end{scope}
}

\node[and port, anchor=in 1](AND0) at (0, 0) {};
\draw (AND0.out) -- ++(\nodesep,0) node[and port, anchor=in 2](AND1){} node[midway, below]{$g_1$};
\draw (AND1.out) -- ++(\nodesep,0) node[ffD, anchor=pin 1](D0){} node[midway, above]{$g_2$};

 \draw (D0.pin 6) node[above, anchor = south west ] { ${lo}_1$} |- ++(-5,1)  |- (AND0.in 1) ;

 \draw (AND1.out) node[above, anchor = west] {} to[short, *-] ++(0,-1.5) |- ++(1.75,0) node[ anchor=west](O1){$o_1$};
 \draw (AND0.in 2) node[ anchor=east](IN1){$a$};
 \draw (AND1.in 1) node[ anchor=east](IN2){$b$};

 \seqframe{0}{6.5}{-2};
 \seqframee{1}{12.2}{-2};

 \draw(11.5,0.75)[color=blue, very thick] -- (12.2,0.75);

\node (x)[color=blue] at (6,0.8) {$X$} ;

\node (a) at (2, -2.5) [draw,draw=none] {$\textbf{(a)}$};
\node (b) at (12.1, -2.5) [draw,draw=none] {$\textbf{(b)}$};

\end{tikzpicture}
\caption{A sequential logic network with feedback (a) and its inductive case network (b) for $1$-step sequential induction before applying assumptions in the first frame. }
\label{fig:seq_assumptions}
\end{figure}
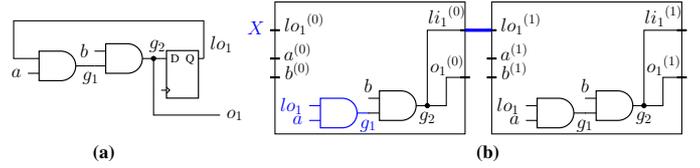

\Cref{fig:seq_assumptions}(a) shows a simple sequential network with feedback whose output is always zero provided that the initial state of the register is zero.
With assumption, our proposed method is able to prove this. For the base case, it is clear that \(g_1\) is stuck at zero.
For the inductive case (shown in \cref{fig:seq_assumptions}(b)), if we assume \(g_1\) of the first frame is stuck at zero, then so is \(g_1\) in the second frame.
This is also an example of a sequential network that is not optimized by retiming based methods.

\subsection{Complete Algorithm}
\label{sec:algorithm}

The high-level pseudocode of our method is presented in \cref{alg:seqrr} including all enhancements.
In \cref{algline:gate}, the algorithm iterates over all gates in the original network, and in \cref{algline:cand}, it considers different optimization candidates \(\Delta\), namely, different fanin redundancies and resubstitutions.
Then, for each frame of the base case network, the algorithm checks if \(\Delta\) is valid and if so, applies \(\Delta\) in that frame (\cref{algline:basecase}-\cref{algline:basecase-end}).
If it is valid in all frames, then it applies \(\Delta\) in the first \(k\) frames of inductive case network (\cref{algline:inductive-k}) and checks for the validity in the last frame (\cref{algline:valid2}).
If valid, it applies \(\Delta\) in the last frame as well and also in the original network (\cref{algline:inductive-last1}-\cref{algline:inductive-last2}).
At any point, if \(\Delta\) is invalid for the considered frame, it undoes all preceding applications of it (\cref{algline:base-undo}, \cref{algline:inductive-undo}).

In \cref{algline:valid1,algline:valid2}, to check for the validity of a target optimization, the algorithm first constructs a window around
the target node in the respective network.
(Note that the window is not restricted to the considered frame; to take the reachable states into consideration, the window should span to all previous frames in general.)
Then it encodes the following as a SAT problem:
\emph{Is there an input pattern (for the window) that would make at least one window output differ if the candidate optimization is applied?}
If the problem is UNSAT, then the target optimization is valid.
The size of the window and the conflict limit for the SAT solver are configurable parameters.

\begin{algorithm}[t]
\small
\DontPrintSemicolon
\SetKwFunction{cost}{cost}
\SetKwInOut{Input}{input}
\SetKwInOut{Output}{output}
\SetKwProg{Fn}{function}{:}{}
\SetKw{KwRet}{return}
\Input{Input network \(N\). Number of frames \(k\).} 
\(N^b \gets \) N unrolled into \(k\) frames and first frame ROs replaced with initial states. \;
\(N^i \gets \) N unrolled into \(k+1\) frames. \;
Let \(N^{b,j}, N^{i,j}\) denote the \(j\)-th frame of \(N^b, N^i\) respectively. \;

\For{each gate \(g \in N\)} 
{
\label{algline:gate}
\For{each candidate optimization \(\Delta\) for \(g\)}  { 
\label{algline:cand}
\For{j = \(1, \dots, k\)}
{
\label{algline:basecase}
\If{ \(\Delta\) is invalid for \(g\) in \(N^{b,j}\)}
{
\label{algline:valid1}
Undo \(\Delta\) in all frames \(N^{b,1}, \dots, N^{b,j-1}\).\;
\label{algline:base-undo}
Continue outer loop.\;
}
Apply \(\Delta\) in \(N^{b,j}\).\;
\label{algline:basecase-end}
}
\label{algline:inductive-k}
Apply \(\Delta\) in \(N^{i,1}, \dots, N^{i,k} \).\;
\If{ \(\Delta\) is invalid for \(g\) in \(N^{i,k+1}\)} 
{
\label{algline:valid2}
Undo \(\Delta\) in \(N^{b,1}, \dots, N^{b,k}\) and \(N^{i,1}, \dots, N^{i,k}\).\; 
\label{algline:inductive-undo}
Continue loop. \;
}
Apply \(\Delta\) in \(N^{i,k+1}\).\;
\label{algline:inductive-last1}
Apply \(\Delta\) in \(N\).\;
\label{algline:inductive-last2}
}
}
return \(N\) \;
\caption{High-level pseudocode of sequential optimization with $k$-step induction and assumptions.}
\label{alg:seqrr}
\end{algorithm}

\subsection{Correctness of the Proposed Approach}

In this section, we show the correctness of our algorithm considering the simple case of 1-step sequential induction with no assumptions.
The proof can easily be extended to the case of k-step induction and to the case with assumptions.

\newcommand{\B}{\mathbb{B}}

\begin{theorem}
\label{thm:induction}
Consider a logic network \(N\) and its 1-step base case and inductive versions \(N^b\) and \(N^i\). Let \(\Delta\) be a logic transformation and let \(N_\Delta, N^b_\Delta, N^i_\Delta\), respectively, be the networks obtained by applying \(\Delta\) to N, to all frames of \(N^b\), and to the last frame of \(N^i\).
If \(N^b\) and \(N^i\), respectively, are combinationally equivalent to \(N^b_\Delta\) and \(N^i_\Delta\), then \(N\) and \(N_\Delta\) are sequentially equivalent.
\end{theorem}

\begin{proof}
Let \(m,n,\ell\) be, respectively, the PI, PO, and register counts of \(N\). 
Let \(s_0 \in \B^\ell\) be the initial state, and let \(S \subseteq \B^\ell\) be the set of all reachable states after the initial clock cycle.
For any \(x \in \B^m\), let \(N^b(x) \in \B^\ell \times \B^m\) be the RI/PO pattern of \(N^b\) on input \(x\).
Similarly, let \(N(x,y)\) and \(N^i(x,y)\) be the RI/PO pattern of \(N\) and \(N^i\) on PI pattern \(x \in \B^m\) and RO pattern \(y \in \B^\ell\).
It is sufficient to show that, for any PI pattern \(x \in \B^m\) and state \(y \in \{s_0\} \cup S\), it holds that \(N(x,y) = N_\Delta(x,y)\).

Note that \(N^b(x) = N(x,s_0)\) and \(N^b_\Delta(x) = N_\Delta(x,s_0)\) by construction of \(N^b\), which implies the claim for \(y = s_0\) because we have \(N^b(x) = N^b_\Delta(x)\) due to combinational equivalence of \(N^b\) and \(N^b_\Delta\).
Now let \(S_i\) be the subset of reachable state of \(N\) after \(i\) clock cycles where \(S_0 = \{s_0\}\). Note that \(S = \cup_{i=1}^{\infty}S_i\).
We show that the claim holds for all \(y \in S_{i+1}\) assuming it holds for all \(y \in \cup_{j=0}^{i} S_j\).
For contrary, suppose that there exists some \(y \in S_{i+1}\) and \(x \in \B^m\) such that \(N(x, y) \neq N_\Delta(x,y)\). 
Let \(x', y'\) be the state and input patterns that led to the state \(y\). Then, if we input \(x', x, y'\) to \(N^i\) and \(N^i_\Delta\), their outputs should differ, which is a contradiction because we assumed they are combinationally equivalent.
\end{proof}

\section{Experimental Results} \label{sec:results}
This section shows experimental results obtained using the proposed approach. Presented are both results for \emph{and-inverter graphs}~(AIGs) and after technology mapping within an industrial flow. The approach was evaluated on a sequential benchmark suite from~\cite{opencores}.

First, we present results for AIGs. In the evaluation baseline, we consider an optimization flow consisting of a state-of-the-art sequential optimization \cite{scorr} together with combinational rewriting (commands \emph{scorr} and \emph{rewrite} in ABC~\cite{abc-paper}). These two are interleaved and run until saturation, i.e., no further reduction is observed. 
In the experimental flow, we additionally run the proposed method on top of the baseline, using 1-step sequential induction. Our setup limits the window size to 50k nodes, the divisor count during resubstitution to 100 nodes, and the TFO of a node to 16 levels. 
The proposed method performs both redundancy removal and resubstitution. We also set a tight control on the level count to prevent increasing it during resubstitution.

The results are presented in \cref{tab:results} where columns `NAND2', `Lev', and `FF' show the number of two-input NAND-gates, combinational logic levels, and flip-flops, respectively. 
The last two columns show the runtime of the experimental flow in seconds and the NAND2 reduction over the baseline.
The experimental flow leads to a 4.1\% average area reduction over the baseline, evaluated as the NAND2 count reduction.
It is worth noting that different trade-offs can be achieved between area and runtime.
For instance, a runtime decrease of ~43\% leads to an average gain of 3.6\%. Larger runtime can lead to more improvements. 
All testcases have been verified using sequential verification (\emph{dsec}) in ABC~\cite{abc-paper}.

In \cref{tab:results2}, we present the results obtained in an industrial flow after the initial synthesis and area-oriented technology mapping. The baseline is an industrial flow that does not use any sequential logic synthesis. Flow1 runs two iterations of sequential SAT-sweeping of a mapped network. Flow2 runs Flow1, followed by the proposed method. 
The results are shown as average improvements over the baseline. Our flow achieves a 18.3\% area reduction, compared to the baseline, and 6.9\% reduction, compared to Flow1 with sequential SAT-sweeping. 
Flow2 leads to a 20\% increase in runtime, compared to Flow1. These results confirm that the two methods are orthogonal and the proposed method finds new optimization opportunities on top of state-of-the-art sequential optimizations (i.e., sequential SAT-sweeping). All benchmarks were equivalence-checked using existing sequential verification tools.

\begin{table}[ht]
\caption{Comparison of the proposed method against the baseline} 
\setlength\tabcolsep{1.2pt}
\begin{tabular}{L{1.6cm} R{1cm} R{0.6cm} R{0.8cm} R{1cm} R{0.6cm} R{0.8cm} R{0.8cm} R{1cm}}\toprule
\multicolumn{1}{l}{}  & 
\multicolumn{2}{c}{\textbf{Baseline}} &
\multicolumn{5}{c}{\textbf{Our Method}}  \\
\cmidrule(){2-4}  \cmidrule(l){5-9} 
\multicolumn{1}{L{1.6cm}}{Name} &	
\multicolumn{1}{R{1cm}}{NAND2}	& 
\multicolumn{1}{R{0.6cm}}{Lev} &	
\multicolumn{1}{R{0.8cm}}{FF}	&  
\multicolumn{1}{R{1cm}}{NAND2} &	 
\multicolumn{1}{R{0.6cm}}{Lev} &	
\multicolumn{1}{R{0.8cm}}{FF} &	
\multicolumn{1}{R{0.9cm}}{Time (s)} &
\multicolumn{1}{R{1cm}}{NAND2
\%}\\
\midrule

aes\_core	&	22026	&	32	&	530	&	21061	&	31	&	530	&	2419.6	&	-4.4	\\
des\_area	&	4611	&	37	&	64	&	4594	&	37	&	64	&	108	&	-0.4	\\
des\_perf	&	77288	&	23	&	8808	&	76053	&	23	&	8808	&	1433.5	&	-1.6	\\
ethernet	&	168	&	13	&	47	&	166	&	13	&	47	&	0.1	&	-1.2	\\
i2c	&	931	&	24	&	126	&	889	&	24	&	126	&	1.5	&	-4.5	\\
mem\_ctrl	&	7097	&	31	&	1050	&	6961	&	31	&	1048	&	41.2	&	-1.9	\\
pci\_bridge32	&	17656	&	32	&	3198	&	17292	&	32	&	3198	&	228.1	&	-2.1	\\
pci\_spoci\_ctrl	&	704	&	20	&	60	&	671	&	20	&	60	&	3.8	&	-4.7	\\
sasc	&	597	&	10	&	117	&	568	&	10	&	117	&	0.3	&	-4.9	\\
simple\_spi	&	779	&	12	&	131	&	772	&	12	&	131	&	0.9	&	-0.9	\\
spi	&	3621	&	31	&	229	&	3583	&	31	&	229	&	118.3	&	-1.1	\\
ss\_pcm	&	464	&	9	&	87	&	399	&	9	&	87	&	0.1	&	-14.0	\\
steppermotor	&	138	&	17	&	25	&	125	&	17	&	25	&	0.1	&	-9.4	\\
systemcaes	&	11106	&	42	&	670	&	11070	&	42	&	670	&	204	&	-0.3	\\
systemcdes	&	2696	&	36	&	190	&	2685	&	34	&	190	&	34.4	&	-0.4	\\
tv80	&	7740	&	58	&	359	&	7396	&	57	&	359	&	310.1	&	-4.4	\\
usb\_funct	&	13910	&	27	&	1722	&	13506	&	26	&	1721	&	61	&	-2.9	\\
usb\_phy	&	457	&	12	&	98	&	403	&	11	&	98	&	0.3	&	-11.8	\\
vga\_lcd	&	89555	&	27	&	17032	&	89294	&	27	&	17032	&	3273.5	&	-0.3	\\
wb\_conmax	&	47026	&	32	&	770	&	40184	&	32	&	770	&	655.1	&	-14.5	\\
wb\_dma	&	3283	&	19	&	521	&	3257	&	19	&	521	&	6.4	&	-0.8	\\
\midrule
Average   & & & & & & & & -4.1 \\
\bottomrule
    \end{tabular}
    \label{tab:results}
\end{table}

\begin{table}[!ht]
\centering
\caption{Results after technology mapping for OpenCores designs}
  \def\tabcolsep{3.6pt}
\begin{tabular}{l c c c c c c }
\toprule
Flow & Combo. Area & Seq. Area & \# Cells & Runtime \\
\midrule
Baseline & 1 & 1 & 1 & - \\ \midrule
Flow1 (SSW) & -12.2\% & -4.8\% & -9.6\% & 1 \\ \midrule
Flow2 (SSW + new method) & -18.3\% & -4.8\% & -14.9\% & +20\% \\ 
\bottomrule
\end{tabular}
\label{tab:results2}
\end{table}
\section{Conclusion} \label{sec:conclusion}

In this work, we propose a new scalable sequential optimization method based on induction where ODCs can be effectively used to find optimization opportunities without limiting don't-cares to be compatible (CODCs).
Our method uses ODCs in the context of windowing while employing SAT-based checks to identify optimization opportunities and it is able to find optimization opportunities that were never found before by prior approaches.
When it comes to runtime, our analysis indicates that most of it is spent on SAT solving. We believe that the runtime can be substantially reduced by invoking the SAT solver only if a counter-example guided simulation fails to disprove a target optimization.

We also remark that the use of our method in a state-of-the-art industrial sequential optimization flow shows promising results. While the observed gains are achieved by naively running the new method after the baseline optimization flow, it remains to be seen where in the flow the new method is the most useful.
We hope that the improved sequential optimization offered by the proposed method will inspire further research.

\balance
\bibliographystyle{IEEEtran}
\bibliography{IEEEabrv,references}

\end{document}